\documentclass[aps,prd,floatfix,superscriptaddress,twocolumn,showpacs]{revtex4}

\usepackage{epsfig}
\usepackage{amsmath}
\usepackage{graphicx,psfrag}
\usepackage{dcolumn}
\usepackage{bm}
\usepackage{slashed}
\usepackage{xcolor}
\usepackage{amsfonts}

\newcommand{\beq}{\begin{equation}}
\newcommand{\eeq}{\end{equation}}
\newcommand{\bea}{\begin{eqnarray}}
\newcommand{\eea}{\end{eqnarray}}
\newcommand{\hf} {\frac{1}{2}}

\newcommand{\nn}{\nonumber\\}

\newcommand\eqn[1]     {Eq.\,(\ref{#1})}

\newcommand\fig[1]     {Fig.\,{\ref{#1}}}

\def\eq#1{(\ref{#1})}
\def\s0#1#2{\mbox{\small{$ \frac{#1}{#2} $}}}
\def\0#1#2{\frac{#1}{#2}}

\def\tu{{\tilde u}}

\def\ord#1{{\cal O}(#1)}

\def\mr#1{{\mathrm{#1}}}

\begin{document}

\title{
Berezinskii-Kosterlitz-Thouless transition and criticality 
of an elliptic deformation of the sine-Gordon model}

\author{N. Defenu}
\affiliation{Institut f\"ur Theoretische Physik, Universit\"at 
Heidelberg, D-69120 Heidelberg, Germany}

\author{V. Bacs\'o}
\affiliation{University of Debrecen, P.O.Box 105, H-4010 Debrecen, Hungary}

\author{I. G. M\'ari\'an}
\affiliation{University of Debrecen, P.O.Box 105, H-4010 Debrecen, Hungary}

\author{I. N\'andori}
\affiliation{MTA-DE Particle Physics Research Group, P.O.Box 51, H-4001 
Debrecen, Hungary}
\affiliation{MTA Atomki, P.O. Box 51, H-4001 Debrecen, Hungary} 

\author{A. Trombettoni} 
\affiliation{CNR-IOM DEMOCRITOS Simulation Center, Via Bonomea 265, 
I-34136 Trieste, Italy}
\affiliation{SISSA and INFN, Sezione di Trieste, 
via Bonomea 265, I-34136 Trieste, Italy}

\begin{abstract} 
We introduce and study the properties of a periodic model interpolating between 
the sine-- and the sinh--Gordon theories in $1+1$ dimensions. 
This model shows the peculiarities, due to the preservation 
of the functional form of their potential 
across RG flows, of the two limiting cases: the sine-Gordon, not having 
conventional order/magnetization at finite temperature, but exhibiting 
Berezinskii-Kosterlitz-Thouless (BKT) transition; and 
the sinh-Gordon, not having a phase transition, but being integrable. 
The considered interpolation, which we term as {\em sn-Gordon} model, 
is performed with potentials written in terms 
of Jacobi functions. The critical 
properties of the sn-Gordon theory are discussed by a renormalization-group
approach. The 
critical points, except the sinh-Gordon one, are found to be 
of BKT type. Explicit 
expressions for the critical coupling as a function of the elliptic modulus 
are given.
\end{abstract}


\maketitle

%-----------------------------------------
\section{Introduction}
\label{sec_intro}
%-----------------------------------------

Symmetries and dimensionality play a crucial role in the 
determination of critical properties 
and phase diagrams. As an example, in quantum field theory one 
of the most studied model 
is the Ising one with interaction terms $\varphi^{4}$ which is known to 
have two phases in $d=1+1$ 
dimensions in one of which the $Z_2$ symmetry has been broken spontaneously \cite{giuseppe}. 
Another paradigmatic and well studied instance of phase transition in $d=2$ dimensions is provided 
by the sine-Gordon (sG) scalar theory where the interaction Lagrangian contains a periodic 
self-interaction $\cos{(\beta \varphi)}$. The sG model has been widely studied for the properties 
of its soliton solutions \cite{drazin,raja} and it  is known to exhibit a Berezinskii-Kosterlitz-Thouless 
(BKT) phase transition \cite{minnhagen87,kadanoff}. Replacing the real valued frequency $\beta$ 
of the sG model by an imaginary one, $\beta \to i \beta$, one arrives at the sinh-Gordon (shG) 
model with a self-interaction term $\cos{(i \beta \varphi)} = \cosh{(\beta \varphi)}$ which is in turn a well 
studied scalar field theory \cite{giuseppe}.

For the shG model the periodicity is lost and no BKT type transition is expected. One could 
argue that, due to its non-periodic nature, the interaction potential can be expanded in Taylor series 
which generates $\varphi^{2N}$ terms, so that one could very 
naively expect an Ising type phase structure. 
However, this is not the case. The shG model is known to possess a single phase, and the explanation 
of this fact is related to the preservation of the functional form of its potential \cite{doreybook}, 
which is connected with the special properties of the exponentials entering the hyperbolic functions. 

Another way to relate the Ising, sG and shG models is based on their conformal properties. 
It is known that systems at criticality, where they are scale-invariant, may give rise to invariance 
under the larger group of conformal transformations \cite{polyakov_1970} locally acting as scale
transformations \cite{di_francesco_1997}. The conformal symmetry in $d=2$ dimensions encloses
  infinitely many local transformations \cite{di_francesco_1997}
  and its occurrence and consequences for $2$-dimensional field theories have been deeply investigated and exploited to 
obtain a variety of exact results \cite{di_francesco_1997,giuseppe}. As a consequence of 
conformal invariance the central charge $c$ is well defined at any fixed point in the phase 
structure of the model and its difference  $\Delta c$ between the one at the Gaussian and 
the non-trivial fixed point characterizes the theory. 
In case of the Ising model, 
$\Delta c \equiv c_\mr{UV} - c_\mr{IR} = 1/2$ where the high-energy (UV) 
value $c_\mr{UV}$ 
is taken at the Gaussian while the low-energy (IR) value $c_\mr{IR}$ is chosen at the 
Wilson-Fisher fixed points. It is known that $\Delta c = 1/2$, 
$1$, $1$ for the Ising, sG and shG models respectively. It is clear that the peculiarities 
of the sG and shG models based on the preservation of the functional form of its potential 
along renormalization group (RG) flows are at the basis of the fact that in both cases 
$\Delta c =1$, with the result for the shG model differing from that of the Ising although it is 
not periodic. %(and it can be considered as a $\varphi^{2N}$ theory).

The goal of the present work is to introduce and discuss a class of models interpolating 
between the sG and the shG models. The critical properties 
of the proposed models can be studied by functional RG, which allows as well to 
clarify from the point of view of the 
interpolation the characteristics of the shG model discussed above. The 
used RG technique is 
well suited to undercover the critical properties, 
at least at qualitative level, of field theoretical 
models, since it maintains the full functional form of 
the effective potential under study. This 
property has proven necessary in order to achieve accurate results 
for $O(N)$ field theories 
in any real dimension and with generic non analytic kinetic terms in the 
effective action 
\cite{Defenu2015a,Defenu2015,Defenu2016}.

The interpolation considered in this paper, that we term {\it sn-Gordon} (snG)
model, is based on Jacobi 
functions \cite{byrd}. 
We remind that the definition of the Jacobi functions follows the same line as the $\sin$ and $\cos$ functions, 
but considering the unit ellipse, rather than the unit circle as the geometrical object
to be described. Denoting $x,y$ the two coordinates in the $\mathbb{R}^{2}$ space, 
all the points of an ellipse with eccentricity $m$ can be parametrised by 
%\begin{align}
%x^{2}+(1-m)y^{2}=1,\quad\mathrm{with}\, 0\leq m\leq 1
%\end{align}
\begin{align}
\label{param}
x=r\cos(\theta),\,\,
y=r\sin(\theta),\,\, r&=\frac{1}{\sqrt{1-m^{2}\sin(\theta)^{2}}}
\end{align}
where $\theta$ is the angle in the $x-y$ plane. Starting from this definition one can define the Jacopi amplitude, i.e. the angular arc length of the ellipse
\begin{align}
u(\theta,m)=\int_{0}^{\theta}\frac{d\omega}{1-m^{2}\sin^{2}(\omega)}.
\end{align} 
Rephrasing the relations in Eq.\,\eqref{param} in terms of the two variables $u,m$ and proceeding in analogy with the
 trigonometric case one gets to the following definitions
 \begin{align}
 \label{jfun}
 x=\frac{\mr{cn}(u,m)}{\mr{dn}(u,m)},\quad
 y=\frac{\mr{sn}(u,m)}{\mr{dn}(u,m)},\quad r=\frac{1}{\mr{dn}(u,m)}
 \end{align}
 for the fundamental Jacobi functions.

In order to introduce the considered interpolation we preliminarly observe that 
the Jacobi functions $\mr{sn}(\beta \varphi,m)$, $\mr{cn}(\beta \varphi,m)$
reduces respectively to $\sin(\beta\varphi)$, $\cos(\beta\varphi)$ for vanishing elliptic modulus 
($m=0$) and to $\tanh(\beta\varphi)$,
$1/\cosh(\beta\varphi)$ for $m=1$. Indeed, in the $m\to1$ limit, the eccentricity is unity and the ellipse, described by Eqs.\,\eqref{param}, becomes a parabola. As a consequence the functions in Eq.\,\eqref{jfun} cannot be periodic, since they do not represent a closed curve.  Therefore, 
a simple interpolating model with a potential expressed in terms of Jacobi functions 
can be constructed as
\bea
V_{\mr{snG}}(\varphi) =  u \, \mr{cn}(\beta\varphi,m) = u \, \mr{cd}(\beta \varphi, m) \, \mr{nd}(\beta \varphi, m).
\label{snG}
\eea
where $\mr{nd}(\beta\varphi,m)=1/\mr{dn}(\beta\varphi,m)$ and $\mr{cd}(u,m)=\mr{cn}(u,m)/\mr{dn}(u,m)$.
The snG potential \eq{snG} for $m=0$ reads  $u \cos(\beta \varphi)$, while for 
$m=1$ it is $u \cosh(\beta \varphi)$ reducing to the shG potential $V_{\mr{shG}}$. 
We observe that the interpolating potential is periodic (except for $m=1$) and we therefore do expect 
a BKT transition for $0 \le m <1$.
\begin{figure}[ht] 
\begin{center} 
\epsfig{file=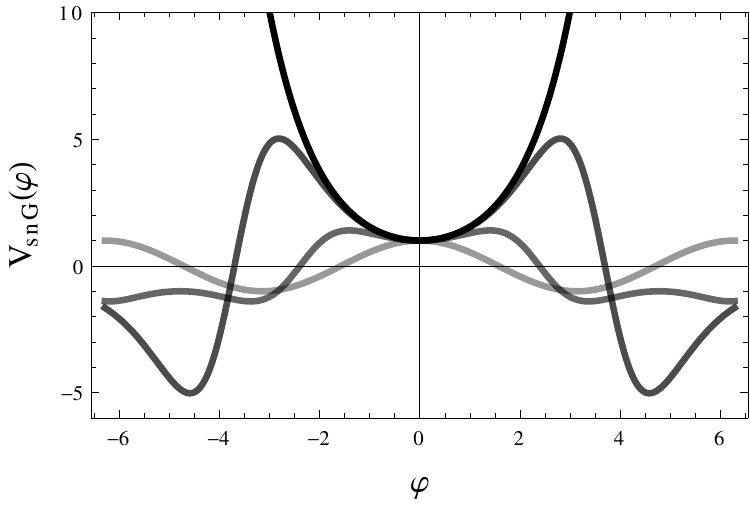,width=8.0 cm}
\caption{
\label{fig0}
The snG potential $V_{\mr{snG}}(\varphi)$ as a function of $\varphi$ for $u=1$ with $m=(0,0.85,0.99,1)$ from bottom to top respectively
while we set $\beta \equiv 1$.
Note that the $m=0$ case (lightest grey curve) is equivalent to the sine-Gordon potential $V_{sG}(\varphi)=u\cos(\varphi)$, while 
the darkest curve for $m=1$ becomes non-periodic and represents the sinh-Gordon potential  $V_{shG}(\varphi)=u\cosh(\varphi)$  }
\end{center}
\end{figure}

An important comment is that, while the sG and the shG models are integrable both 
at classical and quantum level, models interpolating between them are in general
not integrable (we refer to \cite{Olshanetsky1981,Olshanetsky1983} for a discussion of $1+1$ classical and quantum 
integrable models). In the case considered in the present paper, \eq{snG} provides an elliptic 
interpolation between the sG and the shG model, deforming/generalizing the sG potential 
with $m$ being the deformation parameter. We observe that suitable elliptic deformations can be integrable, 
as one can see in one-dimensional non-relativistic systems of classical particles interacting 
via a potential given by a Weierstrass function which reduces to potentials of the form $1/\sin^2(x)$ 
and  $1/\sinh^2(x)$, where $x$ is the distance between the two particles \cite{Olshanetsky1981}. Another example 
is provided by integrable elliptic generalizations of the Calogero model \cite{Langmann2014}. 
In this paper we will not deal with the, 
actually very interesting, problem of constructing integrable elliptic deformations of the sG model and to study 
their soliton-like solutions, but we are primarily interested in introducing a generalization of the sG model 
interpolating between the sG itself and the shG to study the BKT transition across the interpolation 
between these two paradigmatic models. From this point of view the parametrization \eq{snG} represents 
one of the simplest one can think of and suitable to study critical BKT transitions in elliptically deformed models. 

The approach followed here, with the interpolation inserted via the 
potential \eq{snG} in the Lagrangian, is different from the models in which the 
interpolation is done directly in the $S$-matrix, as the staircase model in which an 
analytic continuation of the shG $S$-matrix is performed  to describe interpolating 
flows between minimal models in $2D$ \cite{Zamolodchikov2006}. These interpolating 
models, studied in relation to the so-called ``roaming'',  are integrable by construction. 
In the staircase model a real parameter $\theta_0$ encodes the distance of the continued 
$S$-matrix from the shG self-dual point: in the limit of large $\theta_0$, the ground-state 
energy found by thermodynamic Bethe ansatz exhibits a sequence of scaling behaviours 
approximating those of the minimal conformal field theories. Several aspects of staircase 
and related models were studied 
\cite{martins92,dorey93,ahn1993,delfino95,delfino96,Dorey2015,Horvath16}, including 
a study of the form factors of the shG field \cite{fring} when the real parameter $\theta_0$ 
is sent to infinity \cite{ahn1993} (see more references in \cite{Horvath16}). In these 
models one typically does not work with the Lagrangian (and to reconstruct the Lagrangian 
corresponding to their $S$-matrices is not straightforward) -- at variance the model with the 
snG potential \eq{snG} defines a bare Lagrangian, but anyway one can ask the fate of RG 
flow in the interpolation between sG and shG models.

Finally let us mention an example of interpolation 
done at the Lagrangian level by considering the coupling constant $\beta=\beta_1+i\beta_2$ 
as a complex quantity, where $\beta_1$ and $\beta_2$ are real value frequencies. We note 
that the resulting class of theories can be treated for each non-zero $\beta_2$ as a scalar 
polynomial field theory and denoted as as the Shine-Gordon model. Since we are 
interested mainly in the study of BKT universality class we stick to model \eq{snG}, while 
model with $\beta=\beta_1+i\beta_2$ could be studied in relation to the roaming phenomena. 
Finally, we observe that a 
$(2+1)$ integrable interpolation between the sG and the shG models has been already 
proposed in \cite{Pempinelli1987}, while in brane-world gravity context a modification
of the shG model has been considered \cite{Mannheim2005}.

A disclaimer here, before going {\em in medias res}, is certainly due. 
As mentioned in \cite{raja}, the convention of denoting 
the generalization of the Klein-Gordon model to sinusoidal potential as ``sine-Gordon'' 
generated a certain amount of controversy. If from this point of view the proliferation of 
similar abbreviations should be avoided, from the other the use of sine-Gordon and 
sinh-Gordon models has become so widespread both in physics and mathematics literature 
that in this paper devoted to Lagrangian 
interpolation between these two limits we decided for the purpose of compactness to refer 
to the model \eq{snG} as sn-Gordon. 

The paper is organized as follows. Section II is devoted to introduce the functional RG formalism for 
the study of the snG model. We also discuss there the linearized RG equations. 
In Section III we discuss in detail the limiting cases of the snG 
corresponding to shG and sG. We use functional RG to
discuss also how the standard results are retrieved in these two cases, including the 
point that the shG model does not have a phase transition and the subtleties of the $m \to 1$ limit. 
The discussion of Section III provides the basis for our main 
results exposed in Section IV where we give the functional RG treatment of 
the snG and we discuss the critical properties and the critical values of 
the coupling $\beta$ as a function of the deformation parameter, the 
elliptic modulus m. Section V is devoted to our conclusions.

%-----------------------------------------
\section{Linearized RG equations for the sn-Gordon model}
\label{sec_rg}
%-----------------------------------------

In this section we briefly summarize the functional RG approach for scalar models, and its application 
to the shG and the snG models. 

The functional RG equation has the following form \cite{We1993,Morris94,Berges02,Polony04,Delamotte12}
\beq
\label{RG}
k \partial_k \Gamma_k [\varphi] = \hf \mr{Tr}  
\left[ \frac{k \partial_k R_k}{\Gamma^{(2)}_k [\varphi] + R_k} \right]
\eeq 
for the effective action $\Gamma_k [\varphi]$. $\Gamma^{(2)}_k [\varphi]$ 
denotes the second functional derivative of the effective action and the trace 
$\mr{Tr}$ stands for the integration over all momenta. The RG equation \eq{RG} 
is a functional equation, that should be handled by truncations. Truncated RG flows 
depend on the choice of the regulator function $R_k$, i.e. on the renormalization 
scheme. Regulator functions have already been discussed in the literature by 
introducing its dimensionless form
\beq
R_k( p) = p^2 r(y),
\hskip 0.5cm
y=p^2/k^2,
\nonumber
\eeq
where $r(y)$ is dimensionless. Various types of regulator functions can be chosen, 
but a general choice is the so called CSS regulator \cite{css,css_sg} which recovers 
all major types of regulators in appropriate limits: the Litim \cite{opt_rg}, the power-law 
\cite{Mo1994} and the exponential \cite{We1993} ones. The mass cutoff is the 
power-law regulator $r(y) = y^{-b}$ with $b=1$.

We observe that we do not include the wavefunction renormalization 
in the definition of the regulator when using truncations beyond the leading order of 
the derivative expansion (see below). In this case, in order to ensure scale-invariance 
one has to use the power-law regulator. While this is certainly a restriction we take 
this choice in order to be able to rely on previous results, see e.g., \cite{sG_Trun}.

One of the commonly used systematic approximation is the truncated derivative 
expansion where the action is expanded in powers of the derivative of the field,
\beq
\Gamma_k [\varphi] = \int_x \left[V_k(\varphi) 
+ Z_k(\varphi) \hf (\partial_{\mu} \varphi)^2 + ... \right].  
\nonumber
\eeq 
In the so called Local Potential Approximation (LPA), 
higher derivative terms are neglected and the wave-function renormalization 
is set equal to constant, i.e. $Z_k \equiv 1$. In this case \eq{RG} reduces to the partial 
differential equation for the dimensionless blocked potential ($\tilde V_k = k^{-2} V_k$) 
which has the following form for $d=2$ dimensions 
\bea
\label{lpa}
(2+k\partial_k) {\tilde V}_k(\varphi) = - \frac{1}{4\pi} \int_0^\infty dy 
\frac{y^2 \frac{dr}{dy}}{(1+r)y + {\tilde V''}_k(\varphi)},
\eea
where ${\tilde V''}_k(\varphi)$ is the second derivative of the potential 
with respect to the field.

Before going into the details of the solution of the exact functional RG equation, 
in this section we take the linearized form (around the Gaussian fixed point) 
of the equation \eq{lpa} obtained in the LPA level which reads as
\bea
\label{lpa_lin}
(2+k\partial_k) {\tilde V}_k(\varphi) = - \frac{1}{4\pi} {\tilde V''}_k(\varphi)  + \ord{\tilde V''^2_k},
\eea
independently of the choice of the regulator functions $r(y)$ and apply it to the Ising,
sG, shG and to the interpolating snG models. 

For periodic models which undergo a BKT type phase transition the linearised RG equation at LPA 
can be used to determine the {\em exact} value of the critical frequency $\beta^2_c$ 
which separates the phases of the model. This is a unique property of sG type models
based on the fact that (i) the "critical" fixed point where $\beta^2_c$ is calculated situates
at vanishing Fourier amplitude, (ii) the RG flow equation obtained for the wavefunction 
renormalization beyond the linearised and LPA levels (see e.g.,  Eq.~\eq{single_b1_exact}) 
has {\em no} linear dependence on the Fourier amplitude, thus, for small amplitudes 
it has no scale-dependence at all, hence the exact critical frequency can be obtained 
by the LPA linearised RG equation \eq{lpa_lin}.

%-------------------------------
\subsection{The Ising model}
%-------------------------------

Although it is not the goal of the present work to consider 
the functional RG study of the Ising 
model, since it is useful in the following let us 
first apply \eq{lpa_lin} for the Ising model by substituting  
\beq
\label{ising}
\tilde V_{\mr{Ising}} (\varphi) = \sum_{n=1}^{\mr{N_{CUT}}} \frac{\tilde g_{2n} (k)}{(2n)!} \varphi^{2n},
\eeq
into Eq.~\eq{lpa_lin}. One can then read the RG flow equations for the scale 
dependent dimensionless couplings $\tilde g_{2n}(k)$. 
For any finite $\mr{N_{CUT}}$, the 
linearized functional RG equation does not preserve the functional 
form of the bare theory \eq{ising}, 
i.e., the l.h.s of \eq{lpa_lin} contains polynomial terms $\varphi^{2n}$ 
of order $n=\mr{N_{CUT}}$. The r.h.s of \eq{lpa_lin} has terms of 
order $n<\mr{N_{CUT}}$. Let us note
that the same holds for the case where the linearization 
of the functional RG equation \eq{lpa}
is performed in terms of the field-dependent part of $V''_{k}(\varphi)$ which results
in a regulator-dependent linearized functional RG equation.

%-------------------------------
\subsection{The sG model}
%-------------------------------
The situation is different for the sG model where the bare potential  is defined by
(for the sake of simplicity keeping only the fundamental Fourier mode)
\beq
\label{sg}
\tilde V_{\mr{sG}} (\varphi) = \tilde u_k \cos(\beta \varphi),
\eeq
where the dimensionless Fourier amplitude carries the scale-dependence since
in LPA the frequency $\beta$ does not depend on the running momentum cutoff $k$.
It is clear that the linearized functional RG equation \eq{lpa_lin} preserve the functional form
of the bare potential (no higher harmonics are generated):
\bea
\label{lin_sg}
(2+k\partial_k)\tilde u_k\cos(\beta\varphi) =  \frac{1}{4\pi} \beta^2\tilde u_k\cos(\beta\varphi).
\eea
The RG flow equation for the Fourier amplitude reads
\bea
\label{lin_sg2}
k\partial_k\tilde u_k = \tilde u_k \left(-2+\frac{1}{4\pi}\beta^2 \right),
\eea
with a solution 
\bea
\tilde u_k = \tilde u_{\Lambda} 
\left(\frac{k}{\Lambda}\right)^{-2+\frac{\beta^2}{4\pi}}
\eea
which determines the critical frequency $\beta_c^2 = 8\pi$, where the model undergoes
a BKT-type phase transition \cite{coleman}. It is important to note that even if the bare theory 
of the sG model contains higher harmonics, the linearized functional RG equation \eq{lpa_lin} reduces to 
decoupled flow equations for the Fourier amplitudes of various modes.

%-------------------------------
\subsection{The shG model}
%-------------------------------

By using the replacement $\beta \to i \beta$  in Eq.~\eq{sg}, one finds the bare potential for the 
shG model
\beq
\label{shg}
\tilde V_{\mr{shG}} (\varphi) = \tilde u_k\cos(i\beta\varphi) = \tilde u_k \cosh(\beta \varphi)
\eeq
which is inserted into \eq{lpa_lin} preserving again the functional form of the bare potential:
\bea
\label{lin_shg}
(2+k\partial_k)\tilde u_k\cosh(\beta\varphi) =  - \frac{1}{4\pi} \beta^2\tilde u_k\cosh(\beta\varphi).
\eea
The RG flow equation for the Fourier amplitude reads
\bea
\label{lin_shg2}
k\partial_k\tilde u_k = \tilde u_k \left(-2-\frac{1}{4\pi}\beta^2 \right),
\eea
with a solution 
\bea
\tilde u_k = \tilde u_{\Lambda} \left(\frac{k}{\Lambda}\right)^{-2-\frac{\beta^2}{4\pi}},
\eea
showing that in case of $\beta^2=8\pi$ the exponent does not change sign, hence, the 
shG model has no BKT-type phase transition. In other words, the linearized functional RG of the 
shG model can be derived from the sG model by using the replacement $\beta \to i \beta$ 
which results in a sign change of $\beta^2$ and no BKT-type phase transition.

%-------------------------------
\subsection{The snG model}
%-------------------------------
In the snG model, the dimensionless bare potential reads
\bea
\label{sn}
\tilde V_{\mr{snG}}(\varphi) = \tilde A_k \, \mr{cd}(\beta \varphi, m) \, \mr{nd}(\beta \varphi, m), 
\eea
where the amplitude $\tilde A_k$ is scale-dependent. By using the properties of the Jacobi 
functions $\mr{cd}(u,m) = \mr{cn}(u,m)/\mr{dn}(u,m)$ and $\mr{nd}(u,m) = 1/\mr{dn}(u,m)$ 
it can also be written as
\bea
\label{sn2}
\tilde V_{\mr{snG}}(\varphi) = \tilde A_k \, \mr{cn}(\beta \varphi, m) \, [\mr{nd}(\beta \varphi, m)]^2.
\eea
Inserting Eq.~\eq{sn} or Eq.~\eq{sn2} into the linearized functional RG equation \eq{lpa_lin} one 
observes that the functional form is not preserved since the second derivatives of the 
potential has the following form
\bea
\tilde V''_{\mr{snG}}(\varphi) &=&  \beta^2  \tilde A_k  
\frac{\mr{cn}(\beta \varphi, m)}{\mr{dn}(\beta \varphi, m)^4} \nn
&&\left( 6(m-1) + (5-4m) \, \mr{dn}(\beta \varphi, m)^2
\right). \nonumber
\eea
However, it is important to note that the Jacobi function \eq{sn} is a periodic function, so, it can 
be expanded in Fourier series. One has 
\bea
\mr{cn}(u, m) &=& \frac{2\pi}{K\sqrt{m}} \sum_{n=0}^\infty \frac{q^{n+1/2}}{1+q^{2n+1}} 
\cos\left[(2n+1) \frac{\pi u}{2K} \right],  \nn
\mr{nd}(u, m) &=& \frac{\pi}{2K\sqrt{1-m}} \nn
&+& \frac{2\pi}{K\sqrt{1-m}} \sum_{n=1}^\infty \frac{(-1)^n q^{n}}{1+q^{2n}} 
\cos\left[2n \frac{\pi u}{2K} \right],
\nonumber
\eea
where $q = \exp[-\pi K(1-m)/K(m)]$ and $K(m)$ is the quarter period which can be 
expressed by the hypergeometric function
\bea
K = \int_0^{\pi/2} \frac{d\theta}{\sqrt{1-m \, \sin^2(\theta)}} =   \frac{\pi}{2} \,\,
{}_2F_1\left(\hf,\hf,1,m\right) \nonumber.
\eea
It follows then 
\bea
\label{sn3}
\tilde V_{\mr{snG}}(\varphi) =  \sum_{n=1}^\infty \tilde u_n(k) \cos(n\,b\, \varphi), \,\,
b  = \frac{\beta}{{}_2F_1\left(\hf,\hf,1,m\right)}.
\eea
Inserting \eq{sn3} into the linearized functional RG equation \eq{lpa_lin}, one can derive a 
set of uncoupled differential equations for the Fourier modes
\bea
\label{lin_sn}
k\partial_k\tilde u_n(k) = \tilde u_n(k) \left(-2+\frac{1}{4\pi} n^2 b^2 \right).
\eea
Similarly to the sG model the critical frequency corresponds to the fundamental 
mode, i.e., for $n=1$ where one finds $b_c^2 = 8\pi$ and the higher harmonics 
do not modify it \cite{nandori_sg,schemes}. Thus, one can read the $m$-dependence 
of the original frequency
\bea
\label{sn_crit}
\beta_c^2(m) = 8\pi \, \left[{}_2F_1\left(\hf,\hf,1,m\right)\right]^2
\eea
which clearly signals the existence of a BKT-type phase transition if $m \neq 1$. 
In the limit $m \to 0$ one gets back $\beta_c^2 = 8\pi$, while for $m\to 1$ the original 
frequency blows up and the model has a single phase. Thus, the $m=1$ case the snG 
model undergoes no BKT phase transition. 

\begin{figure}[ht] 
\begin{center} 
\epsfig{file=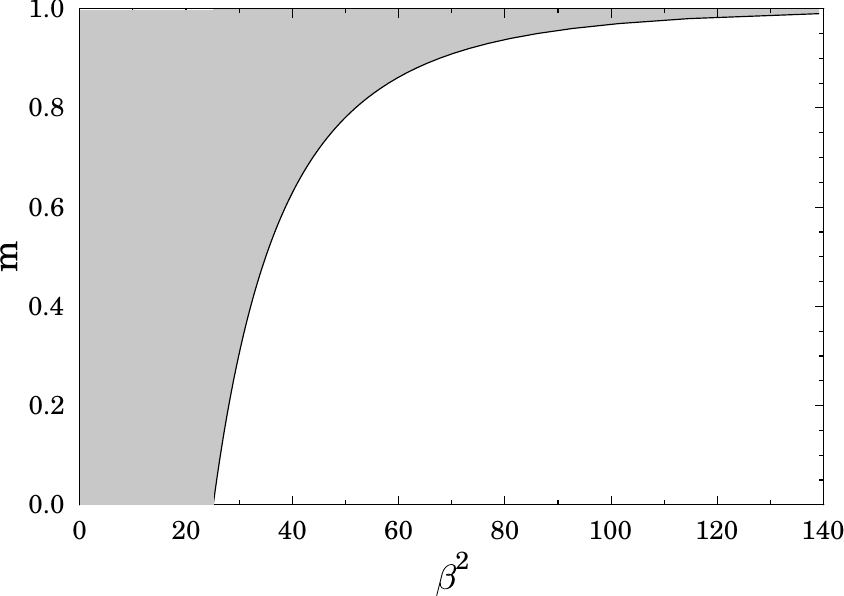,width=8.0 cm}
\caption{
\label{fig1}
Phase structure of the snG model in the $m, \beta^2$ plane based on 
Eq.~\eq{sn_crit} indicating a BKT-type phase transition for $m\neq 1$
where the grey area stands for the massive phase.} 
\end{center}
\end{figure}

However, at this stage we would like to pay the attention of the reader to 
the following important observation. In the limit $m\to 1$ the snG models reduces to 
the shG theory, thus, it is important to study whether the information obtained from the
snG model for $m\to 1$ is in agreement with the results on the shG model. 
The grey area of \fig{fig1} stands for the so called massive phase, where 
the fundamental Fourier amplitude is increasing in the IR limit, so it is expected that
the shG model has to have a single phase with the same properties. We will show in the
next section that indeed, the shG model does not undergo any phase transitions. The 
question which needs to be clarified is whether this single phase of the 
shG model share all features of the massive phase of the snG model and whether to what extent 
the limit $m \to 1$ is singular. We shall come back on the 
limit $m\to 1$ in the next section.

%
% Figure 1
%

In summary, one can conclude that the sG and shG models have a special structure 
such that their functional forms are preserved by the linearized functional RG equation.
A BKT-type phase transition is observed for the sG and the snG models, for the 
latter with a condition $m \neq 1$.

%-----------------------------------------
\section{Functional RG equations} 
\label{sec_exact}
%-----------------------------------------

Here we consider the study of the models introduced in the previous section. 
The functional RG equations are taken in LPA for the Ising model with $\mr{N_{CUT}}=2$ and 
beyond LPA for the other models (keeping only the fundamental mode). 

%----------------------------------------
\subsection{Ising model}
%----------------------------------------
Here we repeat briefly the functional RG study of the Ising model where apart from the trivial mass
term, a $\varphi^4$ self-interaction is taken into account ($N_{\rm CUT} = 2$). The functional RG 
equations are taken in the LPA level, reading in $d=2$ dimensions as  
\bea
\label{g1_g2_mass}
\partial_t g_2 = -2 g_2 - \frac{1}{4\pi} \frac{g_4}{(1+g_2)} \\  
\partial_t g_4 = -2 g_4 +\frac{3}{4\pi} \frac{g_4^2}{(1+g_2)^2}
\eea
for the mass cutoff and 
\bea
\label{g1_g2_litim}
\partial_t g_2 = -2 g_2 - \frac{1}{4\pi} \frac{g_4}{(1+g_2)^2} \\  
\partial_t g_4 = -2 g_4 +\frac{6}{4\pi} \frac{g_4^2}{(1+g_2)^3}
\eea
for the Litim cutoff. The above equations have a trivial Gaussian and a non-trivial 
(cutoff-dependent) Wilson-Fisher (WF) fixed point, where the latter indicates the 
existence of two phases. The $c$-function along the trajectory starting at the Gaussian 
and terminating at the WF fixed points is known to decrease by $\Delta c = 1/2$. 
However, if one consider the massive deformation of the Gaussian fixed point 
$\Delta c = 1$ \cite{c_theorem,c_func}.

%----------------------------------------
\subsection{sG model}
\begin{figure}[ht] 
\begin{center} 
\epsfig{file=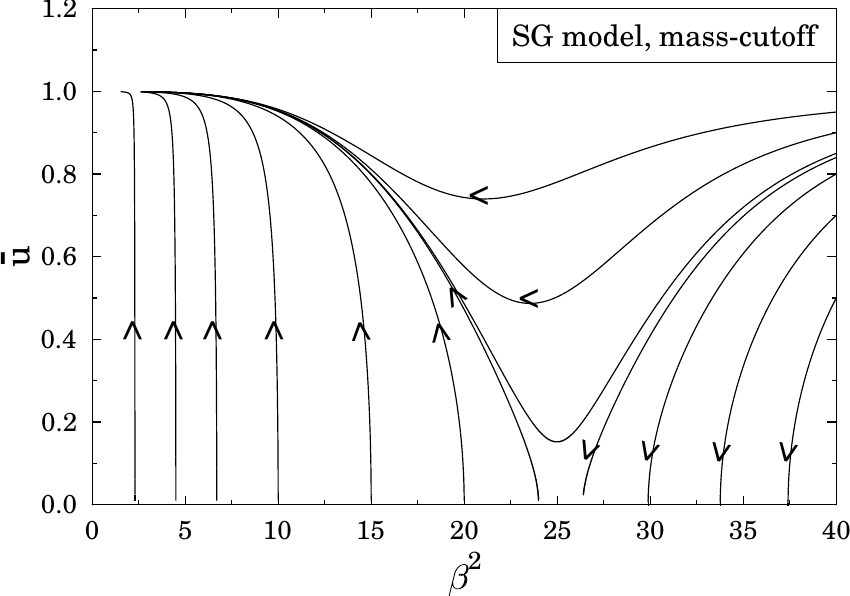,width=8.0 cm}
\caption{
\label{fig2}
Phase structure of the sG model based on Eqs.~\eq{single_b1_exact} indicating 
a BKT-type phase transition with $\beta_c^2 = 8\pi$.} 
\end{center}
\end{figure}

%----------------------------------------
If the sG model \eq{sg} is studied beyond LPA, the RG equation has to be solved 
over the functional subspace spanned by the following ansatz 
\beq
\label{eaans}
\Gamma_k = \int d^2x \left[\frac{1}{2} z_k (\partial_\mu\varphi)^2+V_k(\varphi)\right],
\eeq
where the local potential contains a single Fourier mode 
\beq
\label{z_lpa_sg}
V_k(\varphi) =  - u_k \cos(\varphi),
\eeq
and the following notation is introduced
\beq
\label{identifications}
z_k \equiv 1/\beta^2
\eeq
via the rescaling of the field $\varphi \to \varphi/\beta$ in \eq{sg}, with $z_k$ the 
field-independent wave-function renormalization. Then \eqn{RG} leads to the evolution 
equations for the coupling constants \cite{sG_Trun}, 
\bea
\label{general_ea_u}
k\partial_k u_k &=&
\frac1{2\pi} \int_p ~\frac{ (k\partial_k R_k)}{u_k}\left(\frac{P_k}{\sqrt{P_k^2-u_k^2}}-1\right),\\
\label{general_ea_z}
k\partial_k z_k &=& \frac1{2\pi}\int_p  (k\partial_k R_k)
\biggl(\frac{u_k^2 p^2 (\partial_{p^2}P_k)^2(4P_k^2+u_k^2)}{4(P_k^2-u_k^2)^{7/2}}\nn
&&-\frac{u_k^2P_k(\partial_{p^2}P_k+p^2\partial_{p^2}^2P_k)}{2(P_k^2-u_k^2)^{5/2}} \biggr)
\eea
with $P_k = z_k p^2 + R_k$. In general, the momentum integrals have to be performed 
numerically, however, in some cases analytical results are available. Indeed, by 
using the mass cutoff, i.e. power-law type regulator with $b=1$,  the momentum integrals 
can be performed and the RG equations reads as, 
\bea
\label{single_b1_exact}
(2+k\partial_k)\tu_k &=& \frac{1}{2\pi z_k \tu_k}  \left[1 -  \sqrt{1 - \tu_k^2} \right] \nn
k\partial_k z_k &=& -\frac{1}{24\pi} \frac{\tu_k^2}{[1 - \tu_k^2]^\frac{3}{2}}
\eea
with the dimensionless coupling $\tu = k^{-2} u$. The phase structure of the sG model
based on Eqs.~\eq{single_b1_exact} is plotted on \fig{fig2} which indicates two phases 
with a critical value for the frequency $\beta_c^2 = 8\pi$.

%
% Figure 2
%
Let us note that the power-law regulator with $b=1$, i.e., the mass cutoff has poor 
convergence properties (RG trajectories does not reach the IR fixed point in the 
weak coupling phase), but its advantage that the momentum integral 
can be calculated analytically. A better result can be obtained by using for
example $b=2$, as shown in \cite{c_func_sg}.

%----------------------------------------
\subsection{shG model}
\begin{figure}[ht] 
\begin{center} 
\epsfig{file=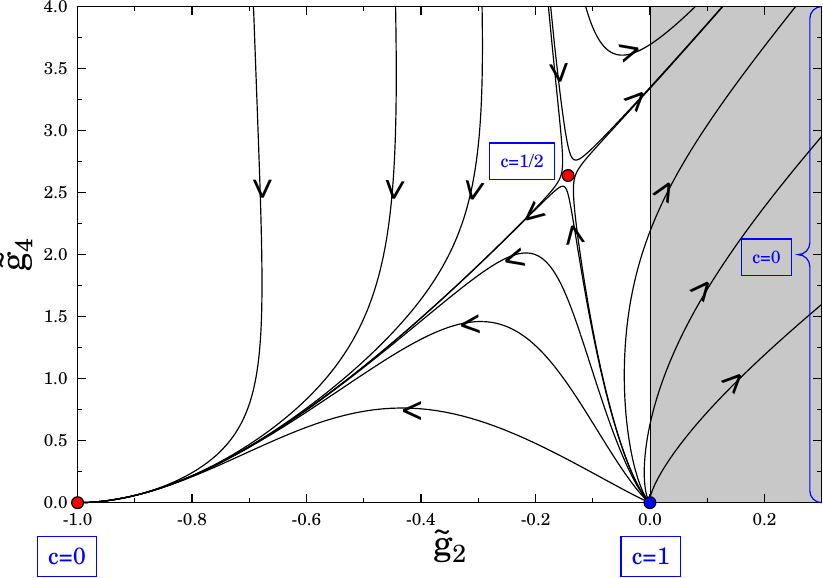,width=8.0 cm}
\caption{
\label{fig3}
Representation of the shG model in the $g_2, g_4$ 
plane based on its Taylor expansion \eq{shg_expand}.
The shaded area stands for initial conditions for the shG model which 
indicates a single phase.} 
\end{center}
\end{figure}

%----------------------------------------
It is important to note that \eq{shg} has a $Z_2$ symmetry, and that the shG model 
is not periodic. Therefore, in order to study the RG flow of the shG model and 
to map out its phase structure one can use the Taylor-expanded form of Eq.~\eq{shg}
\bea
\label{shg_expand}
\tilde V_k(\varphi) &=& \tilde u_k  \left [1+ \frac{1}{2} \beta^2 \varphi^2 + \frac{1}{4!}\beta^4\varphi^4 + ... \right] \nn
&=& \sum_{n=0}^{\infty} \frac{1}{(2n)!}  g_{2n} \varphi^{2n}, \hskip 0.5cm g_{2n} = \tilde u_k \beta^{2n}.
\eea
Thus, the shG model can be considered as an Ising-type model but with restricted initial 
values for the couplings. The key point is that with shG-type initial values the RG flow 
{\em always} starts from the symmetric phase, see \fig{fig3}. Therefore, the shG model 
has a single phase, so, it does not go through a BKT or other type of phase transitions. 

%
% Figure 3
%

The shG model has a special structure that no $2 \to 2n$ particle production is allowed, 
i.e. the production amplitudes of any $2$ particles decay into $2n$ ones are zero at 
tree-level (and also at 1-loop level) \cite{giuseppe,doreybook}. This special structure of 
the bare Lagrangian of the shG model results in a single phase. 

\begin{figure}[ht] 
\begin{center} 
\epsfig{file=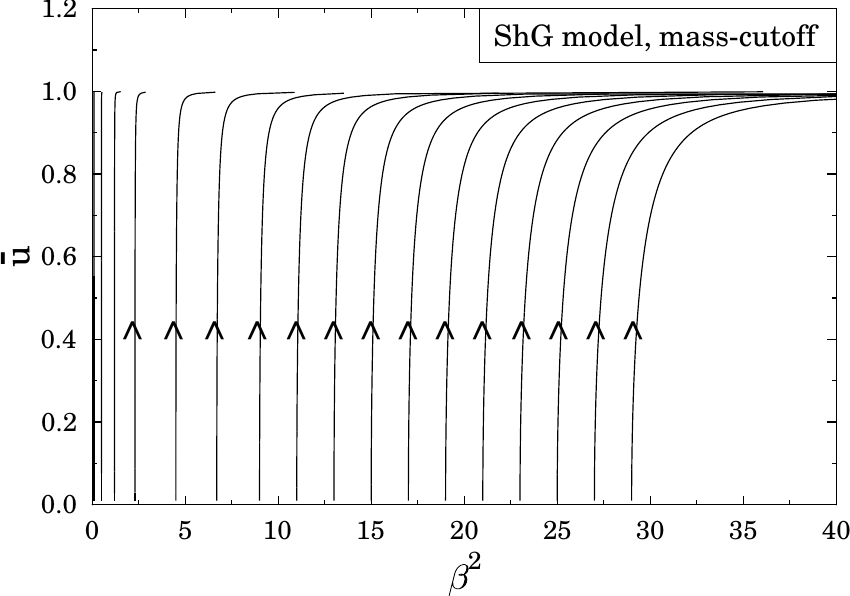,width=8.0 cm}
\caption{
\label{fig4}
Phase structure of the shG model based on \eq{shg_flowa} and 
\eq{shg_flowb} indicates a single phase.} 
\end{center}
\end{figure}

The phase structure of the shG model can also be mapped out by using analytic continuation. 
The simplest way of doing that if one try the replacement of the frequency by an imaginary 
one directly. For example, the RG flow equations for the shG model can be constructed from 
\eq{single_b1_exact} 
\bea
\label{shg_flowa}
(2+k\partial_k) \tilde u_k &=& -\frac{\beta^2}{2\pi\tilde u_k} \left[ 1 - \sqrt{1-\tilde u_k^2} \right] \\
\label{shg_flowb}
k\partial_k\beta_k^2 &=& -\frac{1}{24\pi}\frac{\beta_k^4\tilde u_k^2}{[1-\tilde u_k^2]^\frac{3}{2}}.
\eea
The RG flow of the shG model based on \eq{shg_flowa} and \eq{shg_flowb} is obtained 
numerically and shown in \fig{fig4} which also indicates a single phase for the shG model. 
We observe that due to the poor convergence properties of the regulator ($b=1$ power-law), 
similarly to the sG case, the RG trajectories do not converge properly, specially in the limit 
of vanishing $\beta^2$. 

%
% Figure 4
%

Let us now turn to the study of the $c$-function for the shG model. In our previous paper 
\cite{c_func_sg} we worked out a proper treatment of the $c$-function for the sG scalar 
theory in the framework of functional RG. In the limit of vanishing frequency, the shG and sG 
models become identical to each other, thus the method of \cite{c_func_sg} can be applied 
here for the shG model using the following parametrization 
\bea
\label{shg_scale}
\tilde V_k(\varphi) =  \frac{\tilde m_k^2}{\beta^2} \left(\cos(i \beta\varphi) -1\right),
\eea
where the frequency $\beta$ is assumed to be scale-dependent. In the limit $\beta \to 0$, 
the RG equations for the special form of the shG model \eq{shg_scale} reduce to
\bea
\label{poly_flow_m}
k \partial_k \tilde m_k^2  &\approx& \frac{\tilde m_k^2  [-\beta^2  -8\pi(1+\tilde m_k^2)]}{4\pi (1+\tilde m_k^2)} 
\approx -2 \tilde m_k^2\\
\label{poly_flow_beta}
k \partial_k \beta^2 &=& 0.
\eea
Following the method discussed in \cite{c_func_sg}, the $c$-function of the shG model can 
be determined in the framework of functional RG based on the flow equations \eq{poly_flow_m} and 
\eq{poly_flow_beta} which is identical to that of the sG model in the limit of $\beta\to 0$. Thus, 
the flows for the $c$-function of the shG and the sG models are identical in the limit of vanishing 
frequency, consequently they give us the same result which recovers the known value 
$\Delta c =1$ ($\Delta c = c_{UV}-c_{IR}$) \cite{c_func_sg}.

\begin{figure}[ht] 
\begin{center} 
\epsfig{file=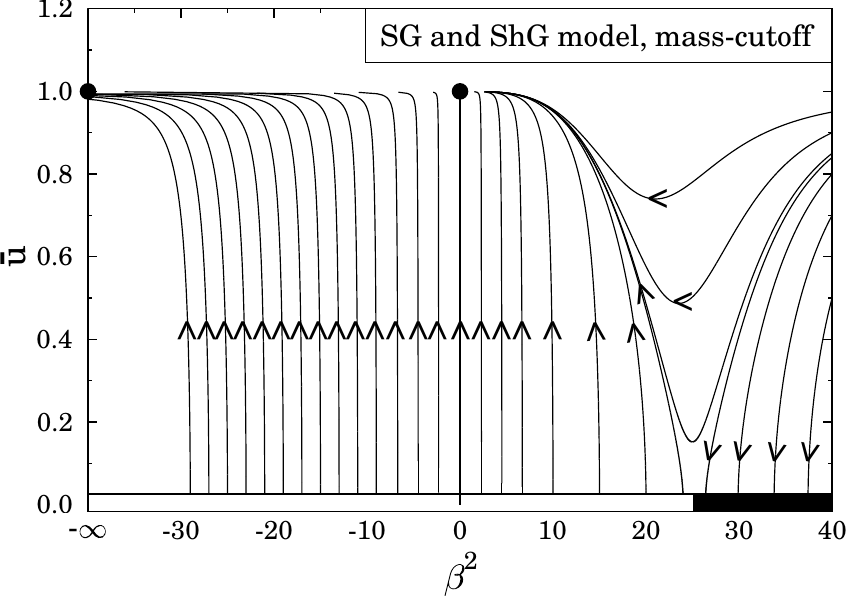,width=8.0 cm}
\caption{
\label{fig5}
Phase structure of the sG model for regions of positive and negative 
$\beta^2$, i.e., the RG flow 
diagram of the sG (\fig{fig2}) and shG (\fig{fig4}) models 
are merged into a single one. Black
circles denote the IR attractive fixed points.} 
\end{center}
\end{figure}

Finally, we briefly discuss on the issue of analytic continuation of the 
sG theory for imaginary frequencies. If one replaces the real value frequency 
by an imaginary one then 
the action of the sG model becomes that of the shG theory. 
This means that one can apply 
the following replacement $\beta^2 \to -\beta^2$ in the flow equations of 
the sG theory  in 
order to obtain the flow equations for the shG model. 
Indeed, the flow diagram of the shG 
model is plotted in \fig{fig4}. This result can be visualised 
in a different way, i.e., by extending 
the sG flow diagram for negative value of the frequency $\beta^2$, 
see \fig{fig5}, which can be compared to figure 1 of Ref.~\cite{malard}. 
There is a disagreement between 
the two figures, namely in \cite{malard} 
the RG trajectories of the negative $\beta^2$ 
regime run into the IR (convexity) fixed point of the sG model which 
signals the presence of spontaneous symmetry breaking (SSB). 
At variance, we argued in this paper that the shG model has no SSB, 
since it has a single phase 
which is the symmetric one. Moreover, 
the flow diagram plotted in figure 1 of \cite{malard}
suggests that the negative and positive $\beta^2$ regions are basically 
reflected to each other, implying in turn the reflection of 
the critical value of the frequency ($\beta_c^2 = 8 \pi$) too. 
However, it was also shown here that no such critical frequency 
exists for the shG model i.e., the 
negative $\beta^2$ case of the sG theory. 
Therefore, we conclude that figure 1 of \cite{malard}
may be misleading and 
we refer to \fig{fig5} below.
%
% Figure 5
%
%

%----------------------------------------
\section{Functional RG study of the snG model}
%----------------------------------------
We are now in position to perform the functional RG study of the snG model. 
According to the previous discussion, it is based on the 
Fourier decomposition \eq{sn3} where the 
frequency $b^2$ of the fundamental mode plays 
a crucial role in the determination of the phase 
structure. Thus, beyond LPA, the snG model can be treated the 
way as the sG model, so the 
RG equation has to be solved over the functional subspace spanned by the following ansatz 
\beq
\Gamma_k = \int d^2x \left[\frac{1}{2} z_k (\partial_\mu\varphi_x)^2+V_k(\varphi_x)\right],
\eeq
where the local potential contains infinitely many Fourier modes 
\beq
\label{z_lpa_sn}
V_k(\varphi) =  - \sum_{n=1}^\infty u_n(k) \cos(n\,\varphi),
\eeq
and the following notations are introduced
\beq
\label{identifications_bis}
z \equiv \frac{1}{b^2} = \frac{\left[{}_2F_1\left(\hf,\hf,1,m\right)\right]^2}{\beta^2}   
\eeq
via the rescaling of the field $\varphi \to \varphi/b$ in \eq{sn3} and $z_k$ again standing for 
the field-independent wave-function renormalization. It is important to note that $m$ remains
a non-scaling parameter even beyond LPA. 

\begin{figure}[ht] 
\begin{center} 
\epsfig{file=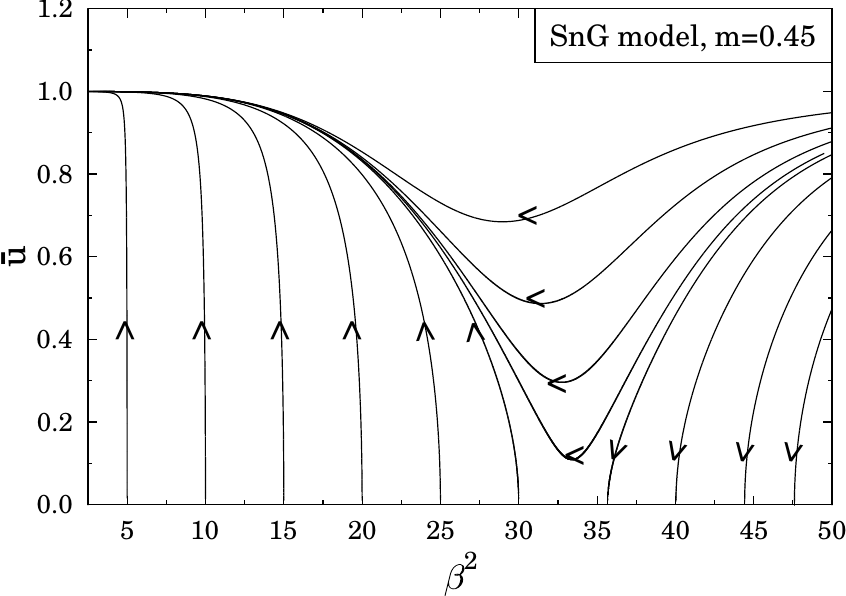,width=8.0 cm}
\caption{
\label{fig6}
Phase structure of the snG model for $m=0.45$, 
indicating a BKT-type phase transition 
with $\beta_c^2 \approx 33.3$.} 
\end{center}
\end{figure}

\begin{figure}[ht] 
\begin{center} 
\epsfig{file=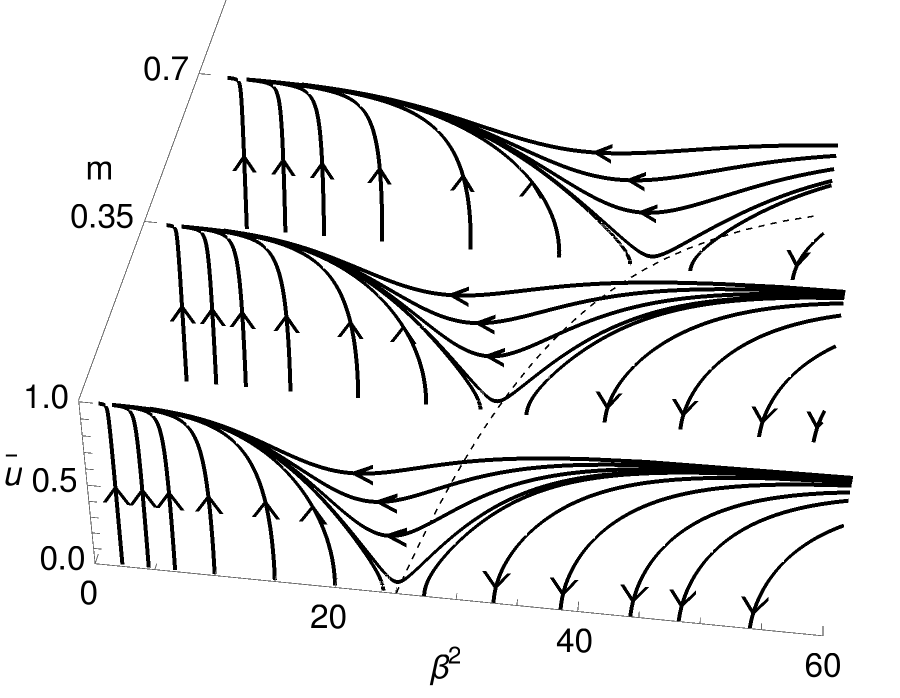,width=8.0 cm}
\caption{
\label{fig7}
Phase structure of the snG model for $m=0, 0.35, 0.7$.
The dashed line indicates the critical frequency $\beta_c^2(m)$ of the BKT phase transition,
similarly to \fig{fig1}.
} 
\end{center}
\end{figure}

In order to follow the strategy done for the sG model one has to take the single-Fourier 
mode approximation of the snG model \eq{z_lpa_sn}. The higher harmonics do not 
change the qualitative picture drawn by the single-Fourier mode approximation 
(for $m\neq 1$) \cite{nandori_sg}. Indeed, by using the mass cutoff, i.e., the 
power-law type regulator with $b=1$, the RG equations for the couplings of the snG reads 
as, 
\bea
\label{single_b1_exact_sn}
(2+k\partial_k)\tu_k &=& \frac{1}{2\pi z_k \tu_k}  \left[1 -  \sqrt{1 - \tu_k^2} \right] \nn
k\partial_k z_k &=& -\frac{1}{24\pi} \frac{\tu_k^2}{[1 - \tu_k^2]^\frac{3}{2}}
\eea
with the dimensionless coupling $\tu = k^{-2} u$ which is identical to the flow equations 
\eq{single_b1_exact} of the sG model but with the different definition for $z$. In order
to compare the flow diagrams of the snG and sG models it is convenient to use the 
squared frequency $\beta^2$ instead of the wave function renormalization $z$. Then, 
the flow diagram of the snG model obtained in the single-Fourier approximation beyond 
LPA for the particular value $m=0.45$ is shown in \fig{fig6}.
%
% Figure 6
%
\fig{fig7} is summarizing our results on the critical properties of snG obtained by RG.
%
%
% Figure 7
%

We finally comment on the limit $m\to 1$ of the snG model. We showed 
that the snG model, being periodic, has a BKT transition in all points but for $m=1$ 
where it reduces to the shG model. Therefore, let us discuss whether the limit 
$m\to 1$ is analytic or not. Two facts that would support the analytic behaviour are following:
(i) the shG as well as the snG model with $m=1$ show a single phase; 
(ii) this phase is the high-temperture one, where the Fourier amplitude is relevant.
However, in favour of the fact that the limit $m\to 1$ is not analytic one can argue 
that (i) the frequency is relevant in the shG model, but irrelevant in the $m\to 1$ limit; and 
(ii) the single phase of the shG model is the symmetric one, but the $m\to 1$ limit suggests SSB.
In order to clearly make a conclusion on the subtleties of the $m \to 1$ limit, 
one has to show a physical quantity which has different value at the two 
cases. To this purpose we propose to the susceptibility of the topological charge
\begin{eqnarray}
\chi=\langle Q^2\rangle-\langle Q\rangle^2
\end{eqnarray}
where $Q$ is the winding number, see \cite{nandori_sg}. This serves as a disorder 
parameter, since the topological susceptibility is vanishing whenever the Fourier amplitude
is zero. This quantity can be shown to be non-zero in the limit $m\to 1$ 
of the snG model, but vanishing for
the shG theory. Therefore, we conclude that the limit $m\to 1$ is non analytic.

%----------------------------------
\section{Summary}
\label{sec_sum}
%----------------------------------

In the present work the renormalization group (RG) study of a class of models 
interpolating between the sine-Gordon (sG) and the sinh-Gordon (shG) theories 
has been addressed. The study of the functional RG equations clearly show 
that only the sG and shG model has a special structure such that their functional 
forms are preserved by the linearized functional RG equations. It was discussed that functional 
RG provides a tool to show that while the sG theory undergoes a phase transition at 
$\beta^2 =8\pi$, this is absent in the shG model.  We argued that the shG model has 
a single phase since it can be considered as an Ising-type model but with restricted 
initial values for the coupling constants. 

We also studied the proposed model, to which we referred as the sn-Gordon (snG) model, 
where the potential is expressed in terms of a product of Jacobi functions. We concluded 
that the snG model exhibits a BKT phase transition for all $m\neq 1$, and we determined 
the phase diagram and the critical value of $\beta$ as a function of the Jacobi parameter 
$m$. These results clearly shows the peculiarities of the two limiting cases, the shG and 
the sG models.

Finally we observe that other interpolations 
between the sG and the shG models can 
be considered. In this paper we focused on the critical properties of 
the snG model, 
but it would be interesting to study also the solitonic solutions 
of the snG model 
and of other possible elliptic interpolations. 
In view of the connection between 
the Ruijsenaars-Schneider models \cite{Rui86} 
and the sG model \cite{Rui86,BabBer}, a deserving investigation would be 
to study the possiblity of integrable interpolations between the sG and the shG 
models.
\\

\section*{Acknowledgement}
The authors gratefully thank
E. Langmann, G. Gori, G. Mussardo, P. Sodano, G. Somogyi, G. Takacs  and T. G. Kovacs
for useful discussions. A.\,T. is grateful for kind hospitality to the 
Galileo Galilei Institute (Florence) where 
part of this work has been performed 
during the Workshop ``From Static to Dynamical Gauge 
Fields with Ultracold Atoms''. 
Financial support by the J\'anos Bolyai Research Scholarship of the Hungarian 
Academy of Sciences and by the CNR/MTA Italy-Hungary 2019-2021 Joint Project 
``Strongly interacting systems in confined geometries" is gratefully acknowledged.
N.\,D. acknowledges financial support by the Deutsche Forschungsgemeinschaft (DFG) under the Collaborative Research Centre ``SFB 1225 ISOQUANT''  and
Germany's Excellence Strategy EXC-2181/1 - 390900948 (the Heidelberg STRUCTURES Excellence Cluster)'.

\end{document}